\documentclass[twocolumn,showpacs,amsmath, preprintnumbers,amssymb, aps, prl]{revtex4-1}
\usepackage{graphicx}
\usepackage{dcolumn}
\usepackage{bm}
\begin{document}
\preprint{deposited by VINITI RAS, 10 December 2014, N335-B2014 }
\title{Casimir EMF }
\author{Evgeny\,G.\,Fateev}
 \email{e.g.fateev@gmail.com}
\affiliation{%
Institute of mechanics, Ural Branch of the RAS, Izhevsk 426067, Russia
}%
\date{\today}
\begin{abstract}
In the present paper, it is shown that the existence of the Casimir 
electromotive force (EMF) is possible in nanosized configurations with 
nonclosed nonparallel metal plates. The nature of such EMF is associated 
with the drag current generation at the noncompensated Casimir action of 
virtual photons on the electrons in the nano-configurations. In the case of 
a classical configuration with strictly parallel plates, EMF is not 
generated. However, EMF can be generated when even an insignificant angle 
between the plates appears. Angles between the plates and their effective 
lengths have been found, at which maximally possible EMF is generated in a 
configuration.
\end{abstract}

\maketitle
\textbf{INTRODUCTION}

Recently, the possibility of the existence of both classical Casimir 
pressure \cite{Casimir:1948, Casimir:1949, Milton:2001, Klimchitskaya:2009, 
Bordag:2009} and expulsive force \cite{Fateev:1, Fateev:2}
in perfectly conducting metal configurations with two 
thin nonparallel plates has been shown. The force shows up as a 
time-constant integral action of virtual photons on the plates (wings) 
forming a configuration with a cavity in the direction of its minimal 
section. In addition, it is shown that periodic configurations with such 
geometry can be expulsed in the same direction under certain conditions 
\cite{Fateev:3}. The total force of the Casimir expulsion of such 
structures should be proportional to the number of geometric entities in the 
configuration. In the present paper, the possibility of the electromotive 
force existence and its expected physical nature are discussed on the basis 
of the above-mentioned nanosized configurations.

It is obvious that no electromotive forces should be generated in parallel 
metal plates in the classical configuration studied by Casimir \cite{Casimir:1948, Casimir:1949} 
However, on the ends of the plates, there can be fluctuations of electric 
potentials due to Johnson-Nyquist thermal noise \cite{Bimonte:2008} and 
electric pick-ups associated with radio interferences. However, when the 
plates in configurations are not parallel, EMF can be generated in such 
systems as it will be shown below. 

\textbf{THEORY}

The EMF nature can be associated with the effects similar to light-induced 
electron drag in metals \cite{Gurevich:1992,Shalaev:1992, Shalaev:1996}, graphite nanofilms 
\cite{Mikheev:2012} and semiconductors \cite{perovich:1981}. As a 
first approximation, the electron-photon drag effect can be explained as 
follows. The momentum of a photon incident at an angle and absorbed by a 
metal surface is transferred to phonons and free electrons in a lattice. In 
this case, the direction of the phonon and electron fluxes does not always 
coincide. However, interacting in the metal, these fluxes will lead to a 
directed electron flux, and thus, to the appearance of EMF at the open 
circuit.

In our case, the perfectly conducting plates experience the action of 
virtual photons having a broad spectral dependence. As a result, these 
photons create the Casimir pressure.
\begin{figure}
\hypertarget{fig1}
\centerline{\includegraphics[width=2in,height=2.2in]{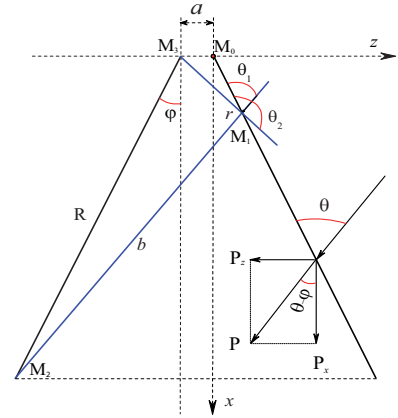}}
\caption{The schematic view of the configuration with nonparallel plates 
(wings) with the wing surface length $R$. The configuration section shown in 
the Cartesian coordinates in the plane $(x,z)$ has the width $L$ in the $y$ 
-direction normal to the plane of the figure. The blue straight lines 
indicate virtual rays with the length $b$ outgoing from the point 
$\mbox{M}_1 $ at the limit angles $\Theta _1 $ and $\Theta _2 $ to the right 
wing and terminating at the ends of the opposite wing of the configuration 
at the points $\mbox{M}_2 $ and $\mbox{M}_3 $, respectively. }
\end{figure}
Let us consider a configuration with nonparallel plates, which will be used 
for the demonstration of the possibility in principle of the Casimir EMF 
existence. The outer and inner surfaces of the wings should have the 
properties of perfect mirrors. The configuration can be immersed into the 
material medium or be its part with the parameters of dielectric 
permittivity different from that of the physical vacuum. The configuration 
in the Cartesian coordinates looks like two thin metal plates with the width 
$L$ (oriented along the $z$ - axis) and surface length $R$; the plates are 
arranged at the distance $a$ from one another and the angle of the opening 
$2\varphi $ between them can be varied by the same value $\varphi $ simultaneously 
for the two wings of the configuration as shown in Fig.\hyperlink{fig1} 1. The particular 
case is parallel planes for $\varphi =0$ and a triangle at $a=0$.

Further, assuming that at any frequency $\omega $, the rays incident onto 
the plate from opposite sides at an arbitrary point $r$ are strictly 
oppositely directed, let us note the following. If we assume that the plates 
are very thin (one atomic layer), the local pressure produced by the 
opposite rays at the point $r$ inside the plate can be added vectorially. In 
this case the virtual ray producing the total pressure $P(r)$ 
\cite{Fateev:1} on the thin plate subsystem will act upon 
electrons. It is clear that the total $P(r)$ will correspond to the 
Casimir pressure at the given point $r$ of the incidence of the ray at the 
angle ${\Theta }'$ to the normal of the wing surface. For finding EMF in the 
entire plate in the configuration it is necessary to find the sum of 
micro-EMF at all the point areas of the plate.

For the first approximation let us use 
the following known formula for the current strength due to the electron 
drag in metals \cite{Goff:1997, Goff:2000}
\begin{equation}
\label{eq1}
\int {d\xi \left\langle {J_{\left\| \right.} } \right\rangle } =\sigma _o 
\frac{\left\langle S \right\rangle }{n_b ec}\left( {1-\rho } \right)\sin 
{\Theta }'\cos {\Theta }'.
\end{equation}
Here $\left\langle S \right\rangle $ is average density of the 
Poynting flux along the incidence of the monochromatic ray at the frequency 
$\omega $, $\sigma _o $ is specific conductivity of metal at constant 
current, $n_b $, $e$ are volume density and electron charge, $\rho $ is 
reflection coefficient, $c$ is light speed, and $\left\langle {J_{\left\| 
\right.} } \right\rangle $ is averaged current generated parallel to the 
plate surface per unit of its thickness $d\xi $. Further, for the ultrathin 
plate of the configuration, let us make a simplifying assumption that
\[
\int {d\xi \left\langle {J_{\left\| \right.} (\xi )} \right\rangle } \approx 
\xi \left\langle {J_{\left\| \right.} } \right\rangle .
\]
The Poynting flux density is expressed through the average density of the 
energy of the electromagnetic wave $\left\langle w \right\rangle $ 
\cite{Ditchburn:1953}
\[
\left\langle w \right\rangle =\frac{1}{c}\left\langle S \right\rangle ,
\]
In this case, the effective pressure on the surface at the mirror reflection 
from it is determined as follows
\begin{equation}
\label{eq2}
P=2\left\langle w \right\rangle =\frac{2}{c}\left\langle S \right\rangle .
\end{equation}
The use of the simple relations of the type (\ref{eq2}) prevents the necessity to 
calculate the Poynting fluxes for virtual electromagnetic waves at all 
possible frequencies $\omega $ because in all the formulae of the Casimir 
pressure all of them are taken into account anyhow. In addition, in the 
presented statement of the EMF problem we confine ourselves to the first 
approximation with no account taken of the possible difference in angles and 
degrees of wave reflection at different frequencies and other possible and 
required further corrections.
\begin{figure*}
\hypertarget{fig2}
\centerline{
\includegraphics[width=1.6in,height=1.6in]{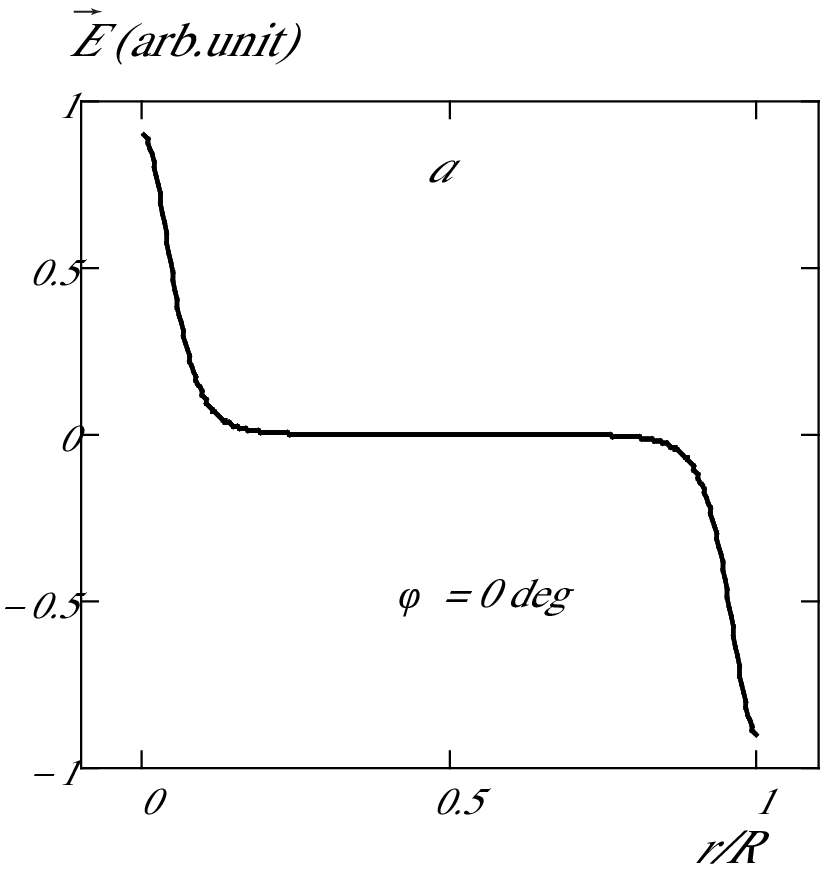}
\includegraphics[width=1.6in,height=1.6in]{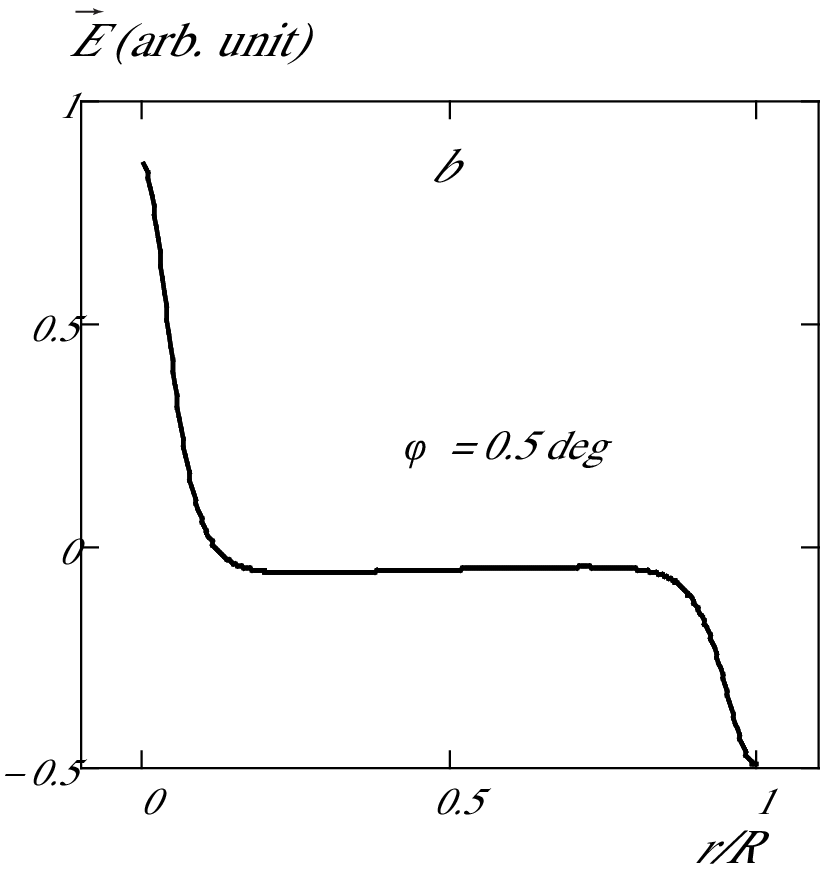}
\includegraphics[width=1.6in,height=1.6in]{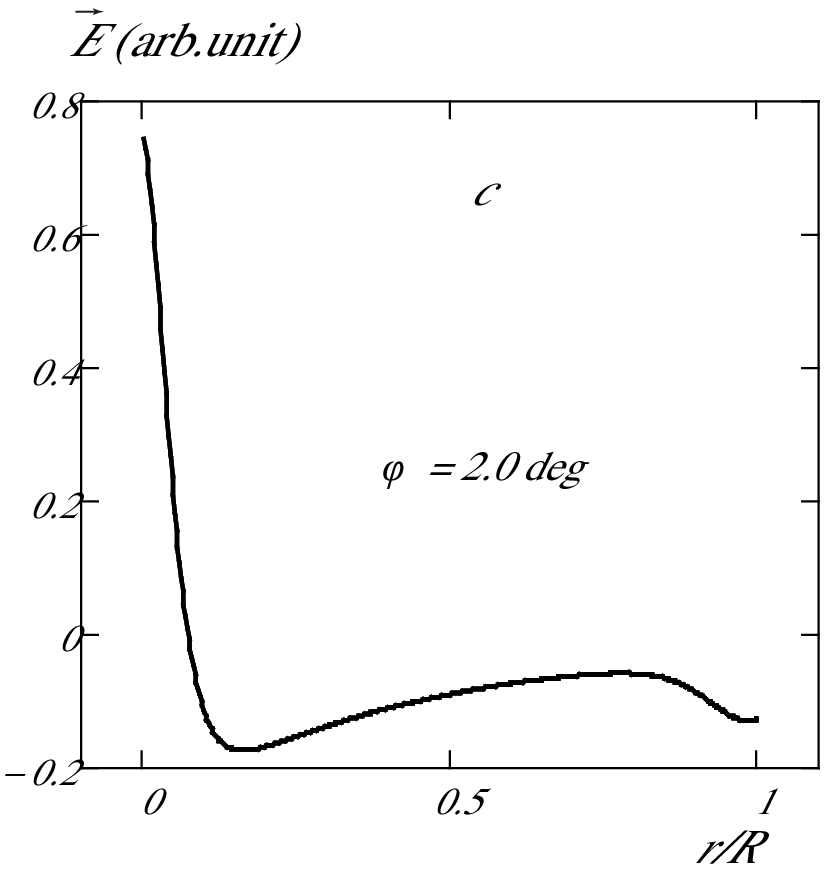}
\includegraphics[width=1.6in,height=1.6in]{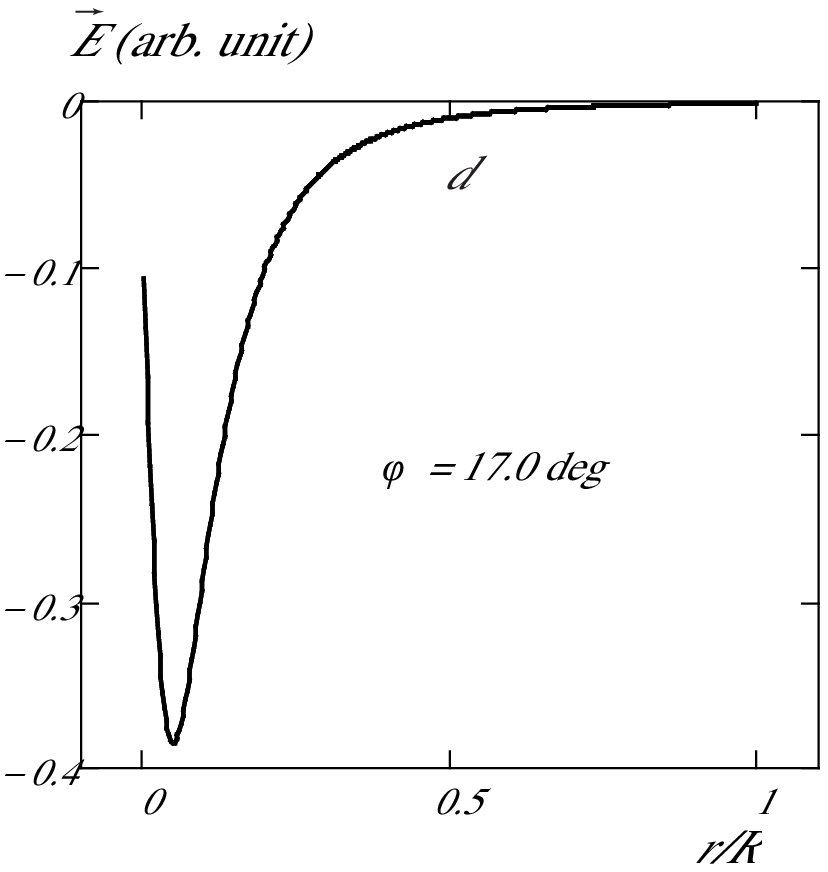}}
\label{fig2}
\caption{The electric field strengths $E(r)$ at local points $r$ on the 
configuration wing with the length $r_{\max } =R$ for four different angles 
$\varphi $.}
\end{figure*}
Thus, taking into account the transmission coefficient of the plate $k$, 
expression (\ref{eq1}) can be written in the form
\begin{equation}
\label{eq3}
\left\langle {J_{\left\| \right.} } \right\rangle =\sigma _o \frac{P(\Theta 
)}{2n_b e \xi}\left[ {1-\rho -k} \right]\sin {\Theta }'\cos {\Theta }'.
\end{equation}
Since electric current is associated with the electric field intensity $E$ 
through the simple relation
\[
\vec {J}_\parallel =\sigma \vec {E}_\parallel ,
\]
at the local point area we find
\begin{equation}
\label{eq4}
\vec {E}_\parallel =\frac{P(\Theta )}{2n_b e \xi}\left[ {1-\rho -k} \right]\sin 
{\Theta }'\cos {\Theta }'.
\end{equation}
Here the total Casimir pressure $P(\Theta )$ at the point $r$ is also 
dependent on the incidence angle $\Theta $ and can be borrowed from 
Ref. \cite{Fateev:1}. Let us note that the $\Theta $ angles for the 
presented geometry of the problem in Fig.\hyperlink{fig1} 1 are determined differently in 
contrast to the ${\Theta }'$ angles for formulae (\ref{eq3}) and (\ref{eq4}) in Refs. 
\cite{Goff:1997, Goff:2000}. As seen from Fig.\hyperlink{fig1} 1, the overdetermination of the angles 
${\Theta }'$ and $\Theta $ obeys the rule
\begin{equation}
\label{eq5}
{\Theta }'=\left\{ {{\begin{array}{*{20}c}
 {\textstyle{\pi \over 2}-\Theta ;\;\Theta <\textstyle{\pi \over 2}} \hfill 
\\
 {\Theta -\textstyle{\pi \over 2};\;\Theta >\textstyle{\pi \over 2}} \hfill 
\\
\end{array} }} \right.,
\end{equation}
and consequently, the trigonometric term in formula (\ref{eq4}) should be rewritten 
as follows
\[
\sin {\Theta }'\cos {\Theta }'=\sin \Theta \cos \Theta .
\]
Thus, EMF can be generated at the point $r$ on each area of the wings 
\begin{equation}
\label{eq6}
\Delta E_\parallel =\int_0^L {dy} \int_{\Theta _1 }^{\Theta _2 } {\vec 
{E}_\parallel dr} .
\end{equation}
The total EMF for the generation areas along the entire length of the wing 
with the length $r_{\max } $ is determined according to the rules of the 
series connection of the sources of EMF
\begin{equation}
\label{eq7}
\Delta E_\parallel =\frac{1}{2n_b e \xi}\left[ {1-\rho -k} \right]\int_0^L {dy} 
\int_0^{r_{\max } } {P(\Theta ,r,\varphi )dr} .
\end{equation}
Here, the local specific pressure $P(\Theta ,r,\varphi )$ at each point $r$ on the 
configuration wing with the length $r_{\max } $and width $L$ has the form 
similar to the expression found in \cite{Fateev:1}; however, the 
term of the type $\cos (\Theta -\varphi )$ responsible for the $x \quad - $and $y \quad - $action 
of the Casimir forces on the tangential component (parallel to the wing 
plane) is replaced by the term of the form $\sin \Theta \cos \Theta $
\begin{equation}
\label{eq8}
\begin{array}{c}
 P(\Theta ,r,\varphi )=-\frac{\hbar c\pi ^2}{240 s^4}\int_{\Theta _1 }^{\Theta 
_2 } {d\Theta } \sin (\Theta -2\varphi )^4\sin \Theta \cos \Theta \\ 
 =-\frac{\hbar c\pi ^2}{240 s^4}A(\varphi ,\Theta _1 ,\Theta _2 ), \\ 
 \end{array}
\end{equation}
\begin{figure*}[htbp]
\hypertarget{fig3}
\centerline{
\includegraphics[width=1.6in,height=1.6in]{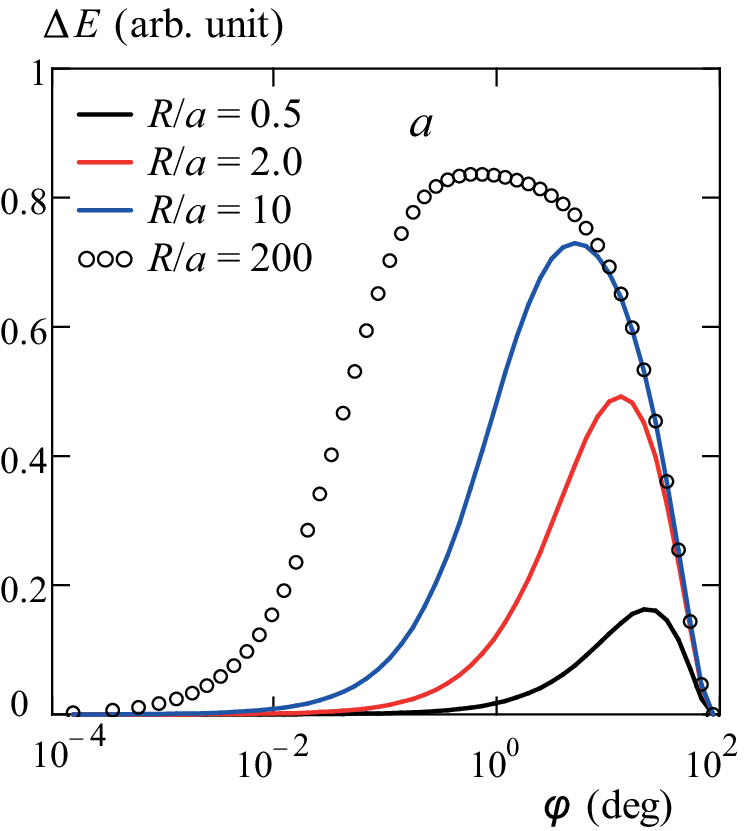}
\includegraphics[width=1.6in,height=1.6in]{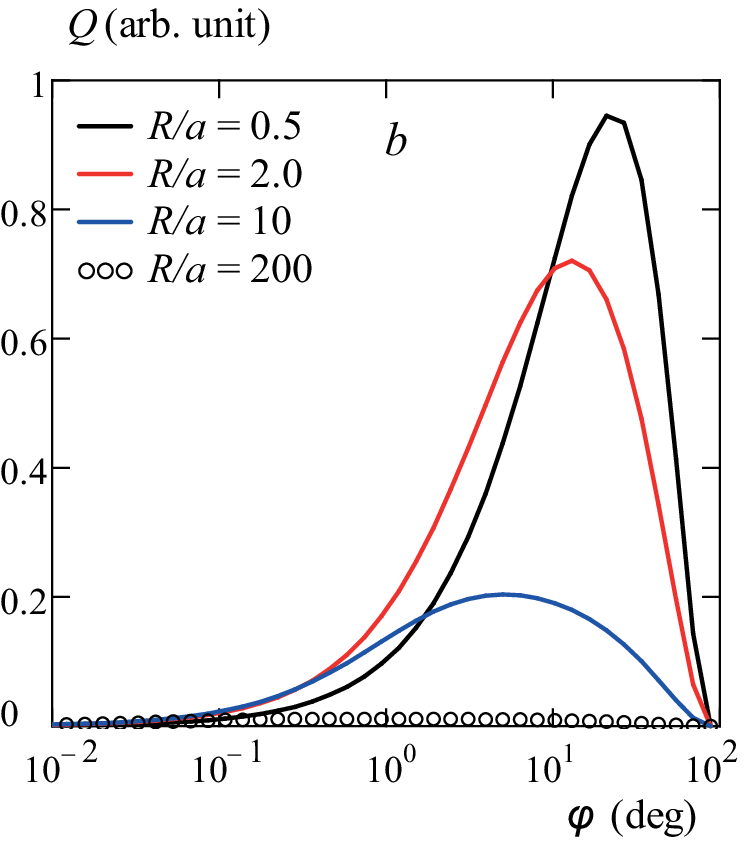}
\includegraphics[width=1.6in,height=1.6in]{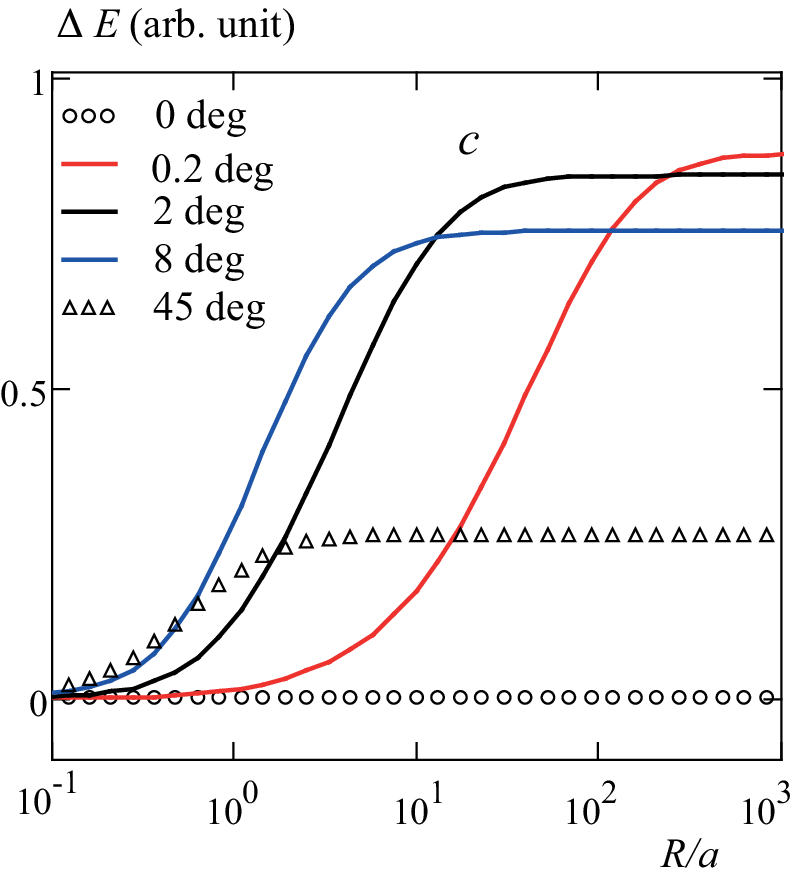}
\includegraphics[width=1.6in,height=1.6in]{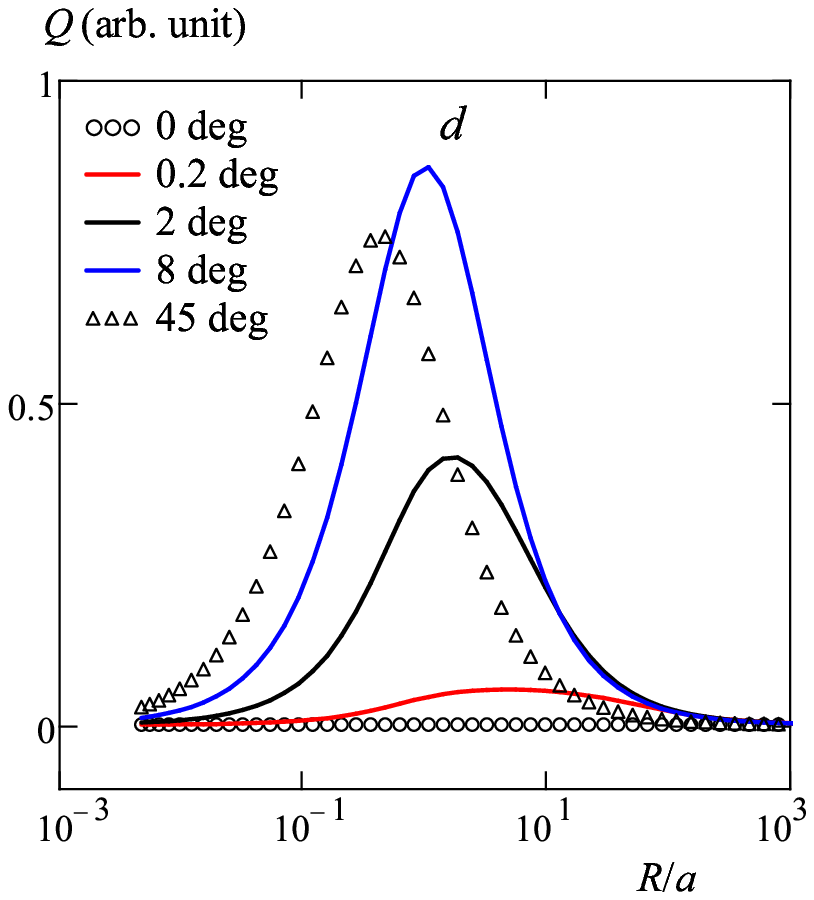}}
\label{fig3}
\caption{($a$) and ($b$) -- the EMF generated in the wing and the effectiveness $Q$ 
depending on the $\varphi $ angles for some lengths $R$. ($c)$ and ($d)$ -- EMF and $Q$ 
effectiveness depending on the wing length $R$ for some $\varphi $ angles. }
\end{figure*}
where
\begin{equation}
\label{eq9}
\begin{array}{c}
 A(\varphi ,\Theta _1 ,\Theta _2 )=\frac{1}{96}\left[ {24\Theta _1 \sin 4\varphi 
-24\Theta _2 \sin 4\varphi } \right. \\ 
 +18\cos 2\Theta _1 -18\cos 2\Theta _2 \\ 
 +6\cos (4\varphi -4\Theta _2 )-6\cos (4\varphi -4\Theta _1 ) \\ 
 +3\cos (8\varphi -2\Theta _2 )-3\cos (8\varphi -2\Theta _1 ) \\ 
 \left. {+\cos (8\varphi -6\Theta _1 )-\cos (8\varphi -6\Theta _2 )} \right]. \\ 
 \end{array}
\end{equation}
In formula (\ref{eq8}), $\hbar =h/2\pi _{ }$ is the reduced Planck constant, $c$ is 
the light speed, and the functional expressions for the limit angles $\Theta 
_1 , \Theta _2 $ in the configuration and the parameter $s$ have the 
forms
\begin{equation}
\label{eq10}
\begin{array}{c}
\Theta _1 =\mbox{arccos}\left[ {-\frac{r+a\sin \varphi -R\cos 2\varphi }{\sqrt 
{\left( {a+R\sin \varphi +r\sin \varphi } \right)^2+\left( {r\cos \varphi -R\cos \varphi 
} \right)^2} }} \right],
 \end{array}
\end{equation}
\begin{equation}
\label{eq10}
\Theta _2 =\mbox{arccos}\left[ {-\frac{r+a\sin \varphi }{\sqrt {a^2+r^2+2ra\sin 
\varphi } }} \right],
\end{equation}
\begin{equation}
\label{eq11}
s=\frac{\sin (2\varphi -\Theta _2 )(a+r\sin \varphi )}{\sin (\varphi -\Theta _2 )}.
\end{equation}
Thus, here a fundamental (idealized) model is presented for the calculation 
of the EMF generation in metal nanosized configurations due to virtual 
photons in optical approximation.

\textbf{RESULTS}

The use of formulae (\ref{eq5}-- \ref{eq11}) allows to reveal the following EMF character 
for two wings with the same length $R$ and minimal distance between them 
$a=4\times 10^{-9}$ m (see Fig.\hyperlink{fig2} 2). The same character of the dependences 
will be observed at the rescaling of the configuration dimensional 
parameters to any values, but naturally, for the ranges restricted by the 
distances between the atoms of well-conducting metals. 

In Fig.\hyperlink{fig2} {2a} it is seen that even when the configuration has parallel plates 
($\varphi =0)$, on the ends of both the right and left wing, electric field is 
generated. The local field strength increases closer to the wing ends and on 
different ends it has opposite direction. In this case, EMF in any of the 
two parallel plates is not generated because of $\Delta E_\parallel 
=E_{_\parallel }^+ +E_{_\parallel }^- =0$. However, at the slightest change 
of the $\varphi $ angle, noncompensation of the local magnitudes $E_{_\parallel 
}^+ $ and $E_{_\parallel }^- $ appears (Fig.\hyperlink{fig2} {2b}). At the further growth of 
the $\varphi $ angle, the noncompensation increases significantly (Fig.\hyperlink{fig2} {2c} and 
\hyperlink{fig2} 2d), which leads to the generation of EMF in the wings, which is the same in 
the direction and value. The current of dragged electrons in both wings is 
directed to the approaching ends of the configuration.

In accordance with formula (\ref{eq5}), the integral quantity of EMF in each of the 
wings depends on the length $r_{\max } =R$ and $\varphi $ angles and has the 
form shown in Fig.\hyperlink{fig3} {3 a, b, c.} 
\begin{figure*}[htbp]
\hypertarget{fig4}
\centerline{
\includegraphics[width=3.4in,height=2.6in]{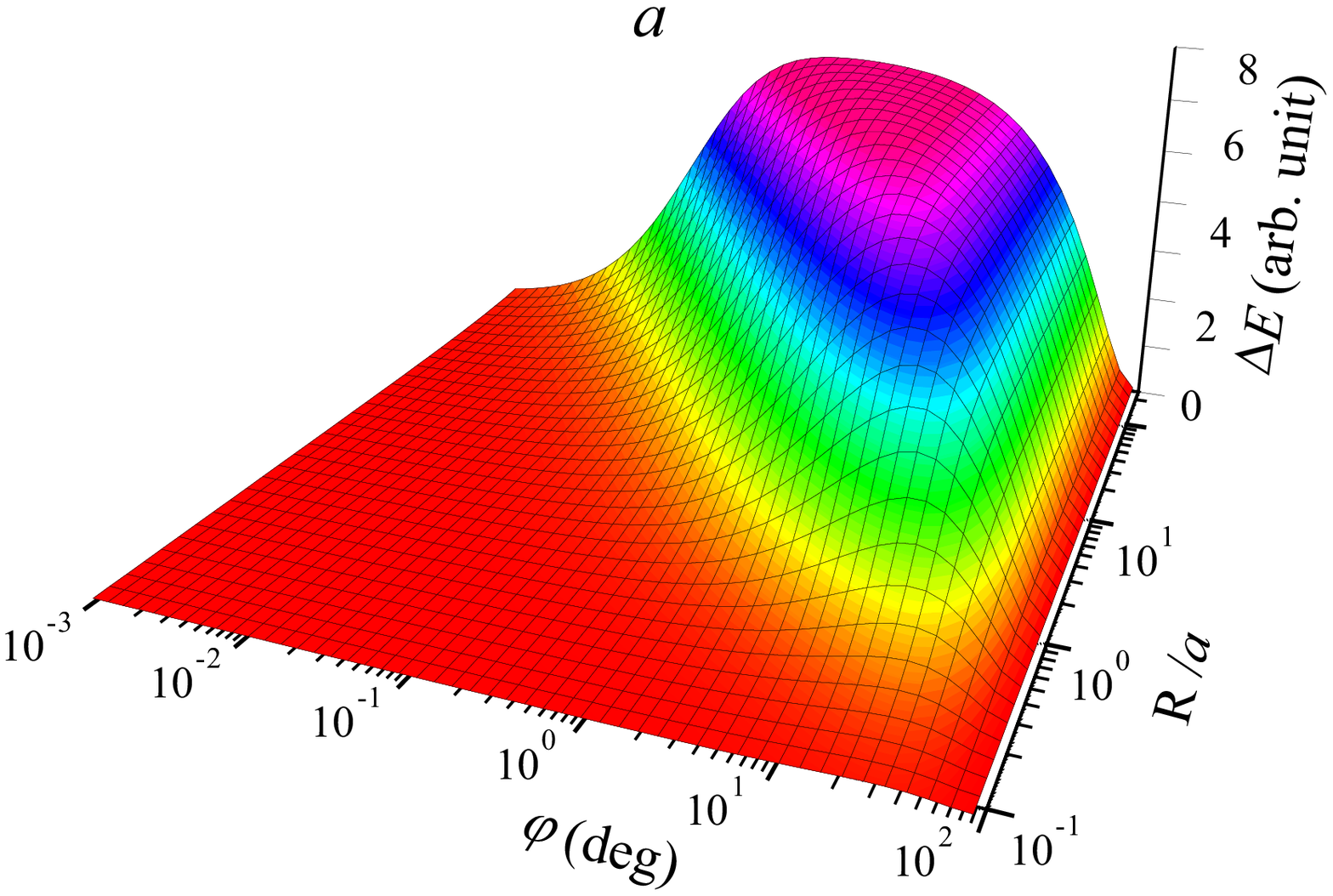}
\includegraphics[width=3.4in,height=2.6in]{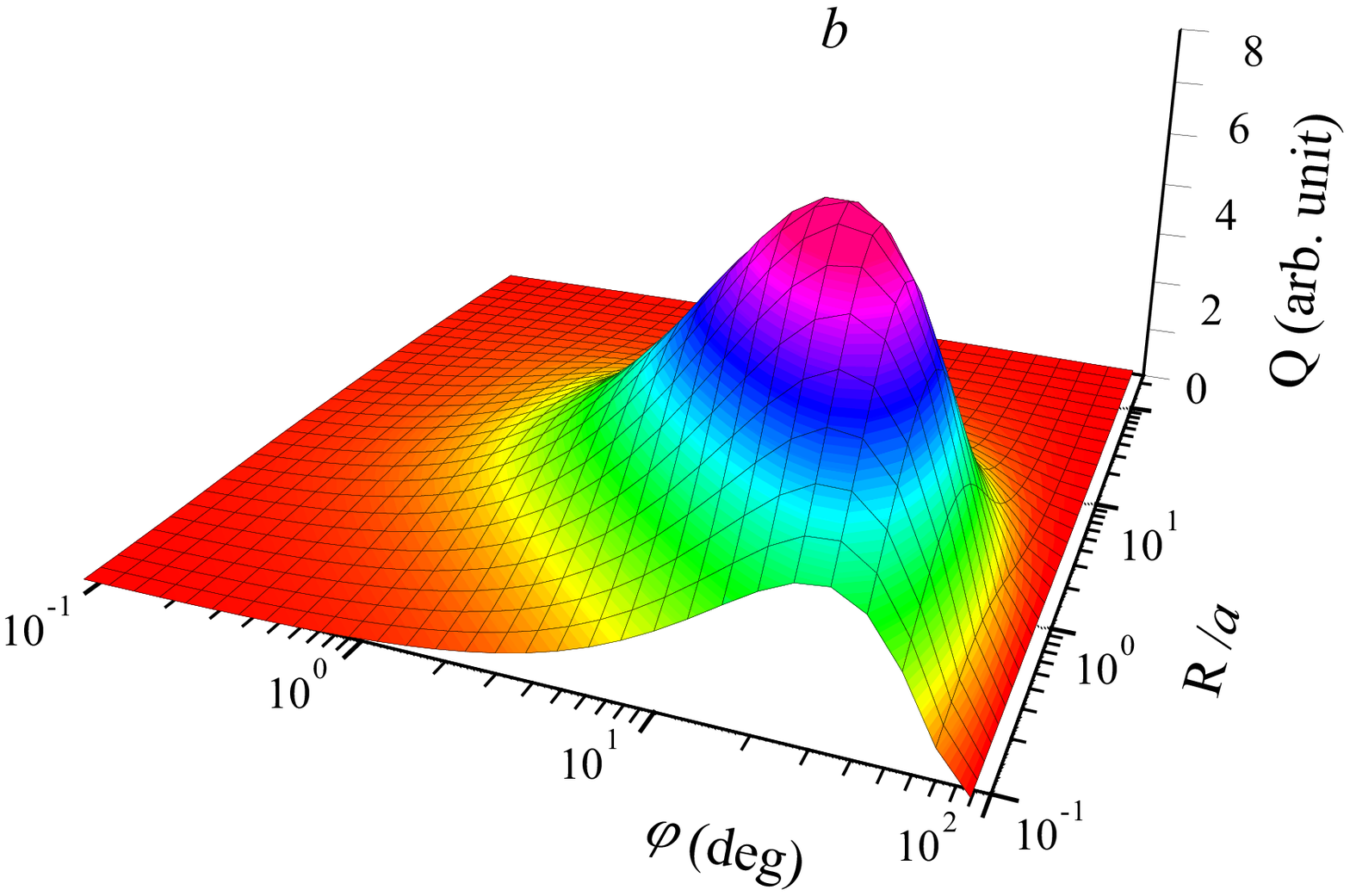}}
\caption{The generated EMF ($a$) in one separate wing of the configuration, and 
the configuration effectiveness (b) $Q$ depending on the $\varphi $ angles and 
wings lengths $R$.}
\end{figure*}
From Fig.\hyperlink{fig3} 3 it follows that at the angle $\varphi =0$, EMF is not generated at 
any wing length $R$. However, at any of the angles $0<\varphi <\pi /2$, EMF is 
generated in any of the wings, and it grows with the increase of the length 
$R$. However, at some values of $R>R_p $, the EMF growth in the wing ceases. 
It means that the main contribution into the EMF generation is made by the 
end areas of the wings, mainly near the ends of the nonparallel wings 
forming the trapezoid configuration. It is clear that there is some optimal 
length of the wings $R$ and optimal angles $\varphi $ (Fig.\hyperlink{fig3} {3b,d}), at which 
there is the maximum of the effectiveness function ratio
\begin{equation}
\label{eq12}
Q=\Delta E/R.
\end{equation}
The effectiveness function maximum shifts with a decrease in the wing length 
$R$ to the large angles $\varphi$ and becomes smaller in its value at the same time. 
All the effects can be observed in two-dimensional dependences $\Delta 
E(R,\varphi )$ and $Q(R,\varphi )$ as it is shown in Fig.\hyperlink{fig4} 4.

Apparently, the excitation of EMF can also be expected in other open 
nanosized configurations. The demonstration of such possibility in two 
closely spaced metal plates is presented as the case which is the simplest 
and most available for calculation. Naturally, when the thickness, 
roughness, temperature and other real parameters of the plates are taken 
into account, the expected value of EMF in the configuration will 
significantly vary.

\textbf{CONLUSION}

Thus, in the present paper, the basic mechanism of the Casimir EMF 
generation in metal open nanosized configurations due to virtual photons has 
been presented. This possibility is theoretically demonstrated using flat 
metal plates (wings) angularly related to one another. In strictly parallel 
plates, EMF is not generated. However, EMF should be generated at any angles 
between the plates in the range of $0<\varphi <\pi /2$ reaching its maximum at 
certain average angles for any wing length. EMF is generated in both wings 
and has similarly situated poles. The optimal values for the wing lengths 
and angles between them have been found, at which the most effective EMF 
generation can take place.
\begin{acknowledgments}
The author is grateful to T. Bakitskaya for hers helpful
participation in discussions.
\end{acknowledgments}

\end{document}